\documentclass[%
 reprint,
 amsmath,amssymb,
 aps,
 prl,
]{revtex4-2}

\usepackage{graphicx}
\usepackage{dcolumn}
\usepackage{bm}

\begin{document}

\preprint{APS/123-QED}

\title{10-GHz-clock time-multiplexed non-degenerate optical parametric oscillator network with a variable planar lightwave circuit interferometer}

\author{Yuya Yonezu}
 \email{yuya.yonezu@ntt.com}
\author{Kensuke Inaba}
\author{Yasuhiro Yamada}
\author{Takuya Ikuta}
\author{Takahiro Inagaki}
\author{Toshimori Honjo}
\author{Hiroki Takesue}
\affiliation{NTT Basic Research Laboratories, NTT Corporation, 3-1, Morinosato Wakamiya, Atsugi, Kanagawa, 243-0198, Japan}






\begin{abstract}
  A coherent XY machine (CXYM) is a physical spin simulator that can simulate the XY model by mapping XY spins onto the continuous phases of non-degenerate optical parametric oscillators (NOPOs). Here, we demonstrated a large-scale CXYM with $>47,000$ spins by generating 10-GHz-clock time-multiplexed NOPO pulses via four-wave mixing in a highly nonlinear fiber inside a fiber ring cavity. By implementing a unidirectional coupling from the $i$-th pulse to the $(i+1)$-th pulse with a variable 1-pulse delay planar lightwave circuit interferometer, we successfully controlled the effective temperature of a one-dimensional XY spin network within two orders of magnitude.
\end{abstract}

\maketitle


Statistical spin models have attracted broad interest due to their potential for solving optimization problems \cite{kirkpatrick1983optimization} as well as for exploring phase transitions and critical phenomena in condensed matter physics \cite{chaikin1995principles}. Recently, Ising machines that can simulate the Ising model by using artificially controllable physical systems have emerged as unconventional computing platforms that offer efficient heuristics for solving computationally-hard problems \cite{mohseni2022ising}. In particular, optical Ising machines, such as coherent Ising machines (CIMs) based on time-multiplexed degenerate optical parametric oscillators (DOPOs) \cite{marandi2014network,takata201616,takesue201610,inagaki2016coherent,mcmahon2016fully,honjo2021100} and spatial-multiplexed photonic Ising machines \cite{pierangeli2019large,babaeian2019single}, have been extensively used to simulate large-scale Ising models by mapping binary spin states onto binary optical phases.

On the other hand, extending the concept of the Ising machine to include simulations of multi-valued spin models (e.g., the Potts model \cite{honari2020optical,inaba2022potts,inoue2023coherent}), continuous-spin models (e.g., the XY model \cite{periwal2021programmable,cosmic2020probing,berloff2017realizing,nixon2013observing,tamate2016simulating,honari2020mapping,takeda2017boltzmann,hamerly2016topological}), or multi-dimensional hyperspins \cite{calvanese2022multidimensional} would be a promising way to increase the applicability of such machines to more general problems. In this regard, XY machines, which simulate the XY model, have been demonstrated in various physical systems with U(1) symmetry, such as atoms \cite{periwal2021programmable}, Josephson junction arrays \cite{cosmic2020probing}, polaritons \cite{berloff2017realizing}, lasers \cite{nixon2013observing,tamate2016simulating,honari2020mapping}, and non-degenerate optical parametric oscillators (NOPOs) \cite{takeda2017boltzmann,hamerly2016topological}. Among these systems, the spatial-multiplexed ones \cite{periwal2021programmable,cosmic2020probing,berloff2017realizing,nixon2013observing} inherently suffer from a difficulty in measuring individual spin states in a large system.

In contrast to the spatial-multiplexed systems \cite{periwal2021programmable,cosmic2020probing,berloff2017realizing,nixon2013observing}, a coherent XY machine (CXYM) based on time-multiplexed NOPOs \cite{takeda2017boltzmann,hamerly2016topological}, which can simulate the XY model by mapping XY spins onto continuous phases, can cope with the scaling up of the system and access all the states and dynamics of each spin because all the phases can be measured with coherent detection techniques \cite{fatadin2008compensation,meyr1998digital}. In addition, the CXYM differs from other time-multiplexed optical simulators \cite{tamate2016simulating} in that the XY spins of the CXYM could be hybridized/interpolated with the Ising spins of the CIM (e.g., ``dimensional annealing'' \cite{calvanese2022multidimensional}) because the CXYM is a complementary machine to the CIM with a similar experimental setup. Recently, Boltzmann sampling of the one-dimensional (1D) XY model was demonstrated with a network of $5,000$ 1-GHz-clock NOPO pulses \cite{takeda2017boltzmann}. The mutual couplings between the NOPO pulses in the nearest-neighbor time slots were implemented with $\pm 1$-pulse ($\pm 1$-ns) delay free-space optical delay lines (ODLs). To construct higher dimensional or complex XY networks, however, scalability and controllability remain significant challenges due to the large footprint of the free-space optics and the need for sophisticated phase control and stabilization for each optical path of the free-space ODLs. Note that experiments on time-multiplexed optical networks using free-space ODLs have also been reported, although the number of the ODLs in those reports was limited to four or less \cite{marandi2014network,takata201616,tamate2016simulating,leefmans2022topological}. 

In this Letter, we describe a demonstration of a large-scale CXYM with $>47,000$ spins by generating 10-GHz-clock time-multiplexed NOPOs and implementing unidirectional interactions with a variable 1-pulse ($0.1$-ns) delay planar lightwave circuit Mach-Zehnder interferometer (PLC-MZI). Increasing the clock frequency (shortening the pulse interval) was found to be beneficial not only for scaling up the number of XY spins but also for stably controlling the interactions between the NOPO pulses because the pulse delay could be implemented with short waveguides integrated in the PLC-MZI. By variating the interactions with variable directional couplers included in the 1-pulse delay PLC-MZI, we successfully controlled effective temperature of the XY spin network within two orders of magnitude. Our PLC-based implementation can be easily extended to more complex networks for the purpose of investigating physics and applications of the XY model and may potentially increase the applicability of unconventional computing platforms based on physical spin simulators.

The dynamics of the NOPO phases $\theta_{i}$ (XY spins) in the CXYM can be described by the Langevin equation \cite{takeda2017boltzmann},
\begin{equation}
  \frac{d\theta_{i}}{dt}=-\frac{\gamma_{\rm{inj}}}{2}\sum \frac{|a_{j}|}{|a_{i}|}J_{ij}\sin \left(\theta_{i}-\theta_{j}\right)+\frac{1}{|a_{i}|}\sqrt{D}\eta_{i}\left(t\right),
  \label{eq:1}
\end{equation}
where $a_{i}=|a_{i}|\exp{(i\theta_{i})}$, $\gamma_{\rm{inj}}$, $J_{ij}$, and $D$ denote the complex amplitude of the $i$-th NOPO pulse, the injection rate, the coupling matrix, and the diffusion coefficient due to noise, respectively. $\eta_{i}\left(t\right)$ is a white-noise function that satisfies $\left<\eta_{k}\left(t_{1}\right)\eta_{l}\left(t_{2}\right)\right>=\delta_{kl}\delta \left(t_{1}-t_{2}\right)$. Assuming that the amplitudes of the NOPO pulses $|a_{i}|$ are all equal in the steady state, $|a_{\rm{SS}}|$, Equation (\ref{eq:1}) represents the noisy Kuramoto model for the case of coupled identical oscillators, which is a non-equilibrium extension of the XY model \cite{rouzaire2021defect}; i.e., the first term of the right-hand side of Eq. (\ref{eq:1}) (the phase drift term) can be rewritten as $-(\gamma_{\rm{inj}}/2)\partial H_{\rm{XY}}/\partial \theta_{i}$ with the XY Hamiltonian $H_{\rm{XY}}=-\sum J_{ij}\cos{\left(\theta_{i}-\theta_{j}\right)}$ ($J_{ij}=J_{ji}$). As a result, the CXYM can simulate the XY model at the effective inverse temperature $\beta J =\gamma_{\rm{inj}}/D_{\theta}$, where $D_{\theta}=D/|a_{\rm{SS}}|^{2}$ is the phase diffusion coefficient, determined by competition between the phase drift due to the interaction and phase diffusion due to noise. Hereafter, we set $J=1$ as a unit of energy.

\begin{figure}[bt!]
  \includegraphics{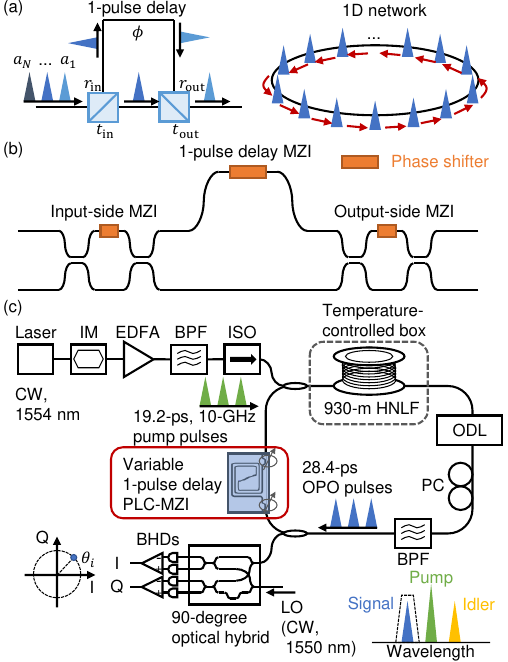}
  \caption{\label{fig:1} Implementation of unidirectional coupling from the $i$-th pulse to the $(i+1)$-th pulse. (a) Schematic diagram of 1-pulse delay interferometer and the implemented 1D network. (b) Variable 1-pulse delay interferometer consisting of three cascaded Mach-Zehnder interferometers (MZIs). (c) Experimental setup. IM: intensity modulator. EDFA: erbium-doped fiber amplifier. BPF: band-pass filter. ISO: optical isolator. HNLF: highly nonlinear fiber. ODL: optical delay line. PC: polarization controller. PLC-MZI: planar lightwave circuit Mach-Zehnder interferometer. LO: local oscillator. BHD: balanced homodyne detector.}
\end{figure}

In this study, we considered a 1D XY network based on the unidirectional coupling from the $i$-th pulse to the $(i+1)$-th pulse implemented with a 1-pulse delay interferometer \cite{hamerly2016topological}, as shown in Fig. \ref{fig:1}(a). The unidirectional coupling violates the symmetric condition of $J_{ij}$ ($J_{ij}\ne J_{ji}$), resulting in a violation of the correspondence between the XY model and steady states, written as $\partial H_{\rm{XY}}/\partial \theta_{i}=0$ with $d\theta_{i}/dt=0$. One of the targets of this study is to clarify the effect of the unidirectional coupling, because the previous studies focused instead on bidirectional Hermite systems \cite{tamate2016simulating,takeda2017boltzmann,leefmans2022topological}.

The input-output relation for the complex amplitude $a_{i}$ of the $i$-th NOPO pulse is given by $a_{i}\rightarrow t_{\rm{out}}t_{\rm{in}}a_{i}+r_{\rm{out}}r_{\rm{in}}e^{i\phi}a_{i-1}$, where $t_{\rm{in}}$, $r_{\rm{in}}$ ($t_{\rm{out}}$, $r_{\rm{out}}$), and $\phi$ represent the amplitude transmission and reflection of the input-side (output-side) beam splitter and the phase difference between the short and long arms of the interferometer, respectively. When the beam splitters are lossless ($t_{\rm{in}}^{2}+r_{\rm{in}}^{2}=t_{\rm{out}}^{2}+r_{\rm{out}}^{2}=1$) and identical ($t_{\rm{in}}=t_{\rm{out}}=\sqrt{T_{\rm{BS}}}$), and $\phi=0$ (ferromagnetic coupling), the input-output relation simplifies to $a_{i}\rightarrow T_{\rm{BS}}a_{i}+\left(1-T_{\rm{BS}}\right)a_{i-1}$. In addition, assuming that the steady-state amplitudes of all NOPO pulses circulating in the cavity are equal to $|a_{\rm{SS}}|$ and the amplitude inhomogeneity is negligible outside of the interferometer, i.e., $|a_{i}|\approx T_{\rm{BS}}|a_{\rm{SS}}|$ and $|a_{i-1}|\approx \left(1-T_{\rm{BS}}\right)|a_{\rm{SS}}|$, Equation (\ref{eq:1}) can be rewritten as
\begin{equation}
  \frac{d\theta_{i}}{dt}=-\frac{\gamma_{\rm{inj}}}{2}\frac{1-T_{\rm{BS}}}{T_{\rm{BS}}}\sin \left(\theta_{i}-\theta_{i-1}\right)+\frac{1}{T_{\rm{BS}}}\sqrt{D_{\theta}}\eta_{i}\left(t\right).
  \label{eq:2}
\end{equation}
The competition between the drift and diffusion terms of the right-hand side of Eq. (\ref{eq:2}) suggests that the dependence of $\beta$ on $T_{\rm{BS}}$ is given by $\beta \propto T_{\rm{BS}}\left(1-T_{\rm{BS}}\right)$. Therefore, $\beta$ can be controlled by adjusting $T_{\rm{BS}}$. To experimentally adjust $T_{\rm{BS}}$, the detailed structure of a variable 1-pulse delay interferometer consists of three cascaded MZIs: an input-side symmetric MZI, an asymmetric MZI including a longer arm for a 1-pulse delay, and an output-side symmetric MZI, as shown in Fig. \ref{fig:1}(b). The phase differences between the arms of the MZIs can be tuned by using thermo-optic phase shifters. Thus, $T_{\rm{BS}}$ can be controlled by variating the phase differences of the input- and output-side symmetric MZIs. 

Figure \ref{fig:1}(c) shows a schematic diagram of the experimental setup. To generate pump pulses, a continuous-wave (CW) laser with a wavelength of 1,554 nm was modulated into 19.2-ps, 10-GHz pulses. The pulses were amplified by an erbium-doped fiber amplifier (EDFA) and then passed through a band-pass filter (BPF) to suppress the amplified spontaneous emission (ASE) noise generated by the EDFA. The amplified pump pulses were injected into a fiber ring cavity, after passing through an optical isolator (ISO). The fiber ring cavity was composed of a highly nonlinear fiber (HNLF) in a temperature-controlled box, a fiber-coupled ODL to adjust the cavity length, a polarization controller (PC), a 0.1-nm-bandwidth BPF with a center wavelength of 1,550 nm, and a variable 1-pulse (0.1-ns) delay PLC-MZI. The phase difference $\phi$ of the PLC-MZI was set to 0 by stabilizing the PLC chip temperature and adjusting the voltage applied to the thermo-optic phase shifter on the long arm.

The NOPO pulses were generated via four-wave mixing in the HNLF, whose nominally specified length, nonlinear parameter, dispersion slope, and zero-dispersion wavelength were 930 m, 21 W$^{-1}$km$^{-1}$, 0.03 nm$^{-2}$km$^{-1}$, and 1,542 nm, respectively. The pump and idler pulses were eliminated through the BPF so that only the signal pulses at a wavelength of 1,550 nm could circulate in the cavity. A portion of the NOPO pulses were extracted from the cavity through a 90:10 coupler to measure the NOPO phases $\theta_{i}$ (XY spins). As shown in Fig. \ref{fig:1}(c), the NOPO phases $\theta_{i}$ were estimated from the separately obtained in-phase and quadrature-phase (IQ) signals of the NOPO pulses by co-injecting the extracted NOPO pulses and an external 1,550-nm CW laser as a local oscillator (LO) into a 90-degree optical hybrid with two balanced homodyne detectors (BHDs). To analyze the raw IQ signals, the IQ imbalance and frequency offset were compensated by using the Gram-Schmidt orthogonalization procedure \cite{fatadin2008compensation} and a frequency estimation via a fast Fourier transform \cite{meyr1998digital}, respectively.

\begin{figure}[bt!]
  \includegraphics{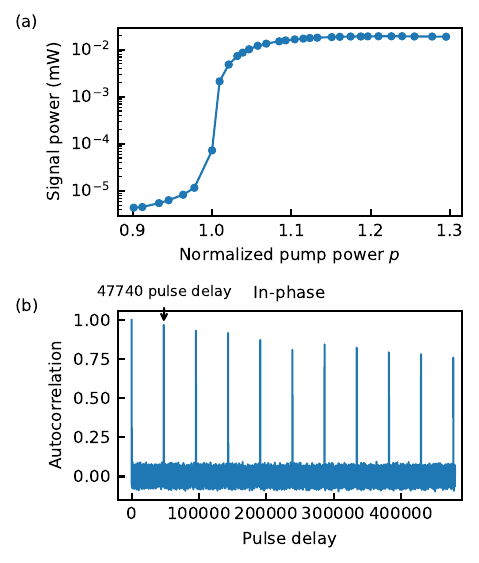}
  \caption{\label{fig:2} Characterization of NOPO pulses at the weakest interaction strength. (a) Monitored NOPO average power as a function of normalized pump power $p$. (b) Autocorrelation of the in-phase signal obtained with the phase measurement using the 90-degree optical hybrid.}
\end{figure}

First, we characterized the generated NOPO pulses at the weakest interaction strength. Figure \ref{fig:2}(a) shows the monitored NOPO signal average power as a function of normalized pump power $p$ (normalized by the threshold pump peak power of $\sim 790$ mW). At the threshold, the NOPO average power changed by more than three orders of magnitude. Figure \ref{fig:2}(b) represents the autocorrelation function calculated from the I signal obtained with the phase measurement of the NOPO pulses at a pump power of $p\sim 1.2$. Clear peaks in the autocorrelation for every additional 47,740 pulses suggest that the phase pattern of the NOPO pulses was preserved for at least ten cavity round-trips. Thus, we confirmed that the CXYM can be scaled up to 47,740 spins. Although in this work the $\sim 1$-km fiber ring cavity was chosen to mitigate optical loss inside the cavity to make it easier to generate the NOPO pulses with a pump power lower than the damage threshold of the BPF, more than 1 million XY spins could be realized by extending the cavity length to $\sim 21$ km, similarly to the previous demonstration for Ising spins based on DOPOs without interactions \cite{takesue201610}.

\begin{figure}[bt!]
  \includegraphics{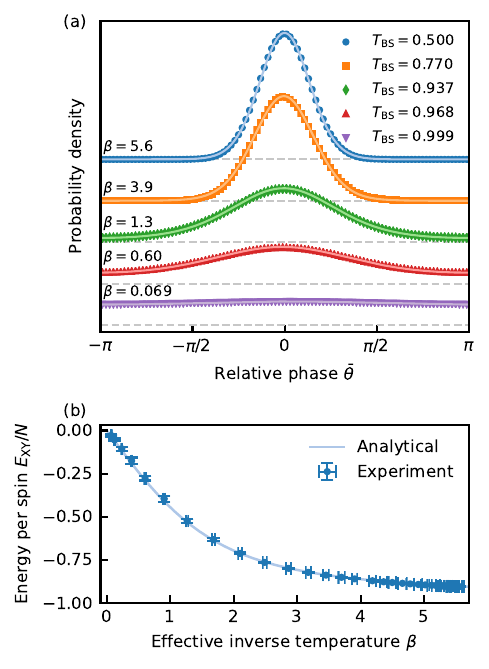}
  \caption{\label{fig:3} Effective temperature control with a variable PLC-MZI. (a) Probability distribution of the relative phases $\bar{\theta}$ between adjacent NOPO pulses. The solid lines represent the theoretical probability distribution for each estimated $\beta$. (b) XY energy per spin as a function of the estimated $\beta$. The error bars indicate the standard deviation. The solid line represents the analytical XY energy for the 1D XY model.}
\end{figure}

To control the effective temperature by tuning the interactions between the NOPO pulses, we changed $T_{\rm{BS}}$ of the input- and output-side symmetric MZIs by using the thermo-optic phase shifters. Figure \ref{fig:3}(a) shows the probability distribution of the relative phases $\bar{\theta}=\theta_{i+1}-\theta_{i}$ between adjacent NOPO pulses obtained from phase measurements for several different $T_{\rm{BS}}$ at a fixed pump power of $p\sim 1.2$. 100 samples of free-running 500,000-NOPO-pulse ($>10$ cavity round-trips) data were acquired for each $T_{\rm{BS}}$. The solid lines in Fig. \ref{fig:3}(a) represent the theoretical curve obtained by fitting the experimental data with the analytical relative phase distribution for the 1D XY model, $P\left(\bar{\theta}\right)=\exp \left[\beta \cos \left(\bar{\theta}+\Delta \theta \right)\right]/2\pi I_{0}\left(\beta \right)$ \cite{tamate2016simulating,takeda2017boltzmann}, where $I_{n}\left(x\right)$ is a modified Bessel function of the first kind. $\beta$ and $\Delta \theta$ are fitting parameters corresponding to the effective inverse temperature and experimental phase error, respectively. Figure \ref{fig:3}(b) shows the dependence of the mean XY energy $E_{\rm{XY}}/N=\left<-\sum_{i=1}^{N}\cos{\left(\theta_{i+1}-\theta_{i}\right)}\right>/N$, where $\left<\ \right>$ represents an average over the acquired samples, on the estimated $\beta$. The solid line in Fig. \ref{fig:3}(b) represents the analytical XY energy of the 1D XY model in the thermodynamic limit, $\lim_{N \to \infty}\left<H_{\rm{XY}}\right>/N=-I_{1}\left(\beta \right)/I_{0}\left(\beta \right)$ \cite{tamate2016simulating,takeda2017boltzmann}. The experimentally obtained XY energy was close to the analytical value of the 1D XY model. These results confirm that the NOPO network reproduced the physics of the 1D XY model even though the unidirectional coupling violates the correspondence between the Hermitian XY model and plausible steady states.

\begin{figure}[bt!]
  \includegraphics{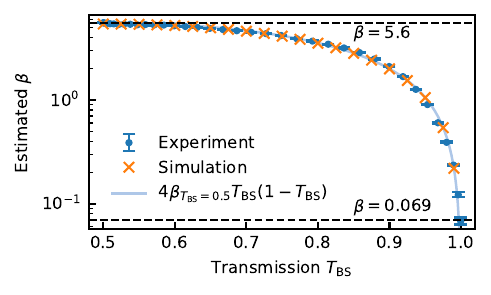}
  \caption{\label{fig:4} Estimated $\beta$ as a function of $T_{\rm{BS}}$ of the input- and output-side variable directional couplers. The error bars indicate the standard deviation. The solid line and cross marks represent $4\beta_{T_{\rm{BS}}=0.5}T_{\rm{BS}}\left(1-T_{\rm{BS}}\right)$ and the numerical results based on Eq. (\ref{eq:2}) with $\gamma_{\rm{inj}}/2D_{\theta}=4\beta_{T_{\rm{BS}}=0.5}$, respectively.}
\end{figure}

Finally, Figure \ref{fig:4} shows the estimated $\beta$ as a function of $T_{\rm{BS}}$. The dependence of $\beta$ on $T_{\rm{BS}}$ agreed well with $4\beta_{T_{\rm{BS}}=0.5}T_{\rm{BS}}\left(1-T_{\rm{BS}}\right)$ (the solid line in Fig. \ref{fig:4}), as suggested from Eq. (\ref{eq:2}). This dependence was confirmed by performing numerical simulations based on Eq. (\ref{eq:2}) with $\gamma_{\rm{inj}}/2D_{\theta}=4\beta_{T_{\rm{BS}}=0.5}$ (the cross marks in Fig. \ref{fig:4}), where the factor $2$ in the denominator of $\gamma_{\rm{inj}}/2D_{\theta}$ originates from the unidirectional coupling. These results indicate that the effective temperature can be controlled within two orders of magnitude ($\beta =0.069\text{--}5.6$). Note that, in the previous study \cite{takeda2017boltzmann}, the effective temperature was controlled in a lower temperature region ($\beta =2.8\text{--}31$) with incoherent noise injection and deviated from the numerical values, while the present system enabled us to accurately control the effective temperature in a wider range thanks to the precise and stable phase controllability and high extinction ratio ($>25$ dB) of the PLC-MZI. 

In summary, we demonstrated a large-scale CXYM with $47,740$ spins based on 10-GHz-clock time-multiplexed NOPOs. We controlled the effective temperature of the 1D XY spin network within the range $\beta =0.069\text{--}5.6$ by tuning the unidirectional nearest-neighbor coupling with a variable 1-pulse delay PLC-MZI. By harnessing the controllability, stability, and compactness of the PLC-MZI, our PLC-based implementation can be used to construct other complex spin network structures, e.g., higher dimensional networks and networks with long-range interactions. For example, the Berezinskii-Kosterlitz-Thouless (BKT) transition in the two-dimensional XY model \cite{hamerly2016topological,rouzaire2021defect,kosterlitz1973ordering}, whose transition temperature is $\beta \approx 1.1$, will be within reach. Thus, the results presented in this paper are a key step towards a useful platform for observing topological phenomena related to the XY model \cite{hamerly2016topological,rouzaire2021defect,kosterlitz1973ordering} as well as for exploring information processing based on XY spins \cite{kalinin2020nonlinear,miri2023neural}.

The authors thank V. M. Bastidas and T. Hatomura for fruitful discussions and H. Tamura for administrative support.


\providecommand{\noopsort}[1]{}\providecommand{\singleletter}[1]{#1}%

\end{document}